\documentstyle[preprint,aps,epsfig]{revtex}
\tightenlines
\begin{document}
\preprint{\vbox{\hbox{IFUM 606/FT \hfill} 
                \hbox{hep-ph/9803390 \hfill}
                \hbox{March 1998 \hfill}}}
\title{ Regge trajectories and quarkonium spectrum from a first
  principle Salpeter equation }
\author{ M. Baldicchi and G.M. Prosperi }
\address{
Dipartimento di Fisica dell'Universit\`{a} di Milano, and\\ 
I.N.F.N., Sezione di Milano, Italy\\ }
\maketitle
\begin{abstract}
We compute the heavy-heavy, light-light and
light-heavy quarkonium spectrum
starting from a first principle Salpeter equation
obtained in a preceding paper.
We neglect spin-orbit structures and exclude from our
treatment the light pseudoscalar states which in principle
would require the use of the full Bethe-Salpeter
equation due to the chiral symmetry breaking problem.
For the rest we find an overall good agreement with the
experimental data. In particular for the light-light case
we find straight Regge trajectories with the right slope
and intercepts.
The strong coupling constant $ \alpha_{\rm s} $, the string
tension $ \sigma $
occurring in the potential and the heavy quark masses
are taken from the heavy quarkonium semirelativistic fit with
only a small rearrangement.
The light quark masses are set equal to baricentral value of
the current quark masses as reported by the particle data group.
For what concerns the light-light and the light-heavy
systems the calculation is essentially parameter free.
\end{abstract}
\setcounter{equation}{0}
\newpage
\section*{}

In reference \cite{prosp96}
a Bethe-Salpeter like equation was derived which provides in principle a
fully relativistic description of the quark-antiquark system.
The derivation takes advantage of an appropriate path integral
representation for a kind of 4-point gauge invariant Green function and
proceeds entirely from first principles apart from the so called
modified area low assumption.
Such assumption consists simply in adding an area term to the
perturbative value of the logarithm of the Wilson loop
correlator, and, as well known, it is well founded
both on lattice simulation and on theoretical considerations
\begin{equation}
  i \ln W = i ( \ln W )_{ \rm{pert} } + \sigma S_{ \rm{min} }.
\label{uno}
\end{equation}

Unfortunately a
direct treatment of the full B-S equation in its four dimensional
form seems to be beyond the present possibilities.
From it, however, by a standard method (which includes replacement
of the full quark propagator by a free propagator
and an instantaneous approximation of the kernel)
a center of mass hamiltonian
can be obtained in the form
\begin{equation}
  H = \sqrt{ m_{1}^{2} + {\bf k}^{2} } +
  \sqrt{ m_{2}^{2} + {\bf k}^{2} } + V,
\label{eq1a}
\end{equation}
with
\begin{eqnarray}
&&  \big< {\bf k} | V | {\bf k'} \big> =
  \frac{1}{2 \sqrt{ w_{1} w_{2}
  w_{1}^{\prime} w_{2}^{\prime} } }
  \bigg\{ \frac{4}{3} \frac{ \alpha_{ \rm{s} } }{\pi^{2}}
  \left[ - \frac{1}{ {\bf Q}^{2} } \left( q_{10}
  q_{20} + {\bf q}^{2} -
  \frac{({\bf Q} \cdot {\bf q})^{2}}{ {\bf Q}^{2} }
  \right) + \right.
\nonumber  \\
&&  + \frac{i}{2 {\bf Q}^{2} } {\bf k'} \times {\bf k} \cdot (
  \mbox{\boldmath $ \sigma $}_{1}
  +
  \mbox{\boldmath $ \sigma $}_{2}
  ) + \frac{1}{2 {\bf Q}^{2}} \bigg[ q_{20} (
  \mbox{\boldmath $ \alpha $}_{1}
  \cdot {\bf Q}) - q_{10} (
  \mbox{\boldmath $ \alpha $}_{2}
  \cdot {\bf Q} ) \bigg] +
\nonumber   \\
&&  +  \frac{1}{6} \left.
  \mbox{\boldmath $ \sigma $}_{1}
  \cdot
  \mbox{\boldmath $ \sigma $}_{2}
  + \frac{1}{4}
  \left( \frac{1}{3}
  \mbox{\boldmath $ \sigma $}_{1}
  \cdot
  \mbox{\boldmath $ \sigma $}_{2} -
  \frac{( {\bf Q} \cdot
  \mbox{\boldmath $ \sigma $}_{1}
  ) ( {\bf Q} \cdot
  \mbox{\boldmath $ \sigma $}_{2}
  )}{ {\bf Q}^{2} }
  \right) +
  \frac{1}{4 {\bf Q}^{2} } (
  \mbox{\boldmath $ \alpha $}_{1}
  \cdot {\bf Q}) (
  \mbox{\boldmath $ \alpha $}_{2}
  \cdot {\bf Q} ) \right] +
\label{due} \\
&&  +  \frac{1}{( 2 \pi )^{3}}
  \int \! d^{3} {\bf r} \, e^{i {\bf Q} \cdot {\bf r} }
  J^{ \rm{inst} }({\bf r},{\bf q},q_{10},q_{20}) \bigg\}
\nonumber
\end{eqnarray}
and
\begin{eqnarray}
&&  J^{ \rm{inst} }({\bf r},{\bf q},q_{10},q_{20}) =
  \frac{ \sigma r }{ q_{10} + q_{20} }
  \bigg[ q_{20}^{2} \sqrt{ q_{10}^{2} - {\bf q}_{ \rm{T} }^{2} } +
  q_{10}^{2} \sqrt{ q_{20}^{2} - {\bf q}_{ \rm{T} }^{2} } +
\nonumber \\
&&  + \frac{ q_{10}^{2} q_{20}^{2} }{ | {\bf q}_{ \rm{T} } | }
  \bigg( \arcsin \frac{ | {\bf q}_{ \rm{T} } | }{q_{10}} +
  \arcsin \frac{ | {\bf q}_{ \rm{T} } | }{q_{20}} \bigg) \bigg]
  - \frac{ \sigma }{ r } \bigg[
  \frac{ q_{20} }{ \sqrt{ q_{10}^{2} - {\bf q}_{ \rm{T} }^{2} } }
  \bigg( {\bf r} \times {\bf q} \cdot
  \mbox{\boldmath $ \sigma $}_{1}
  + i q_{10} ( {\bf r} \cdot
  \mbox{\boldmath $ \alpha $}_{1}
  ) \bigg) +
\nonumber \\
&&  + \frac{ q_{10} }{ \sqrt{ q_{20}^{2} - {\bf q}_{ \rm{T} }^{2} } }
  \bigg( {\bf r} \times {\bf q} \cdot
  \mbox{\boldmath $ \sigma $}_{2}
  - i q_{20} ( {\bf r} \cdot
  \mbox{\boldmath $ \alpha $}_{2}
  ) \bigg) \bigg] .
\label{tre}
\end{eqnarray}

In eq. (\ref{eq1a}-\ref{tre})
the perturbative term has been evaluated only at the first order
in the coupling constant $ \alpha_{\rm s} $,
the indices 1 and 2 denote the quark and the antiquark,
$ {\bf k} $ and $ {\bf k'} $ denote
the final and the initial center of mass
momentum of the quark,
$ \sigma $ is the string tension,
$
  {\bf q} = \frac{ {\bf k} + {\bf k'} }{2}
$;
$
  {\bf Q} = {\bf k'} - {\bf k}
$;
$
  q_{j0} = \frac{ w_{j} + w_{j}^{\prime} }{2},
$
$
  q_{ \rm{T} }^{h} = ( \delta^{h k} - \hat{r}^{h} \hat{r}^{k} ) q^{k}
$
is the transverse momentum, while
$ \alpha_{j}^{k} $ are the usual Dirac matrices $ \gamma_{j}^{0}
\gamma_{j}^{k} $, and $ \sigma_{j}^{k} $ the $ 4 \times 4 $ Dirac spin
matrices
$ i/4 \, \varepsilon^{knm} [ \gamma^{n}_{j} , \gamma^{m}_{j} ] $.

It is shown in \cite{prosp96} that by an
$ 1/m $ expansion and an appropriate
Foldy-Wouthuysen transformation one arrives from
eqs.~(\ref{due}) and~(\ref{tre})
to the semirelativistic potential discussed
in \cite{prspcomo1}.
Such potential is made by a static part $ V_{0} $, a velocity
dependent part $ V_{\rm vd} $ and a spin dependent part
$ V_{\rm sd} $. The expression
$ V_{\rm sd} $ is identical to that derived from the
so called scalar confinement hypothesis while $ V_{\rm vd} $
is different.

As shown in \cite{prspcomo1} and \cite{tstrlt1},
when appropriate values are given
to the quark masses $ m_{j} $, to $ \alpha_{\rm s} $ and to
$ \sigma $, the semirelativistic potential
$ V_{0} + V_{\rm vd} $
reproduces reasonably well the spin averaged 
multiplets in the heavy quarkonium case and it
is phenomenologically favoured with respect to the scalar
confinement expression.
On the other hand it is shown in \cite{olss93} that
$ V_{\rm sd} $ gives the correct fine and hyperfine splittings
if even the second order contributions in $ \alpha_{\rm s} $
are included.

In this paper we want to treat directly the hamiltonian
(\ref{eq1a}-\ref{tre}). In this way we shall see that
we not only reproduce the same results for heavy quarkonia
as from the semirelativistic potential,
but we also obtain the correct ground Regge trajectories
(with the right slope and intercepts) for the systems
involving light quarks alone
and the known lowest states for the systems
with a light quark and a heavy antiquark.
(On the contrary we did not succeed in evaluating daughter
trajectories due to computation difficulties.)
As we shall explain more precisely later, we have achieved this goal
by a very small rearrangement of the parameters used in
\cite{tstrlt1,tstrlt2}
and adopting current masses as given by the particle data group
\cite{prtdatb} for the light quarks $ u $, $ d $ and $ s $.

Notice that to treat light-light and light-heavy systems
various models have been attempted in literature,
based on some conjectural choice of a B-S
kernel (see for instance \cite{durand})
or directly of a local potential in eq. (\ref{eq1a})
(see e.g. \cite{fulcher} and references therein).
However while such models give a reasonable spectrum,
they usually
fail in reproducing the observed Regge trajectories (in
the sense that they do not result straight \cite{durand},
or, if they do, they have not the correct slope and
intercepts) for the values of the string tension
required by the heavy quarkonia fit.

In this paper we shall neglect as a rule the spin
dependent terms in eqs. (\ref{due}) and
(\ref{tre}) and shall mainly pay attention to the
velocity dependent part alone.
Indeed the inclusion of the spin-orbit terms
in the light quark case would make the entire treatment
much more involved due to the impossibility of an expansion
in $ 1/m $ and, as we shall see, it would not be particularly
significant for our analysis.
Later we shall return on S-wave hyperfine splitting
which however would require the inclusion of the
$ \alpha_{\rm s}^{2} $ terms to give quantitative predictions.
Finally in the light-light case we shall restrict our attention to
triplet states alone. In fact for the light pseudoscalar
mesons like $ \pi $, $ K $, $ \eta $, $ \eta^{\prime} $,
the approximation of the quark propagators
by their free expressions as implied in the
use of an hamiltonian is certainly
inadequate, due to the complicate interplay existing in this case
between quark propagator and B-S wave function which is
related to the chiral symmetry breaking \cite{chirale}.

Restricting our consideration
to the velocity dependent part of the potential and
splitting it in a perturbative and in a confinement part,
$ V = V_{ \rm{pert} }  + V_{ \rm{conf} } $, we can write
\begin{equation}
  \big< {\bf k} | V_{ \rm{pert} } | {\bf k'} \big>
  = \frac{1}{2 \sqrt{ w_{1} w_{2}
  w_{1}^{\prime} w_{2}^{\prime} } }
  \left\{ - \frac{4}{3} \frac{ \alpha_{ \rm{s} } }{\pi^{2}}
  \left[ \frac{1}{ {\bf Q}^{2} } \left( q_{10} q_{20} + {\bf q}^{2} -
  \frac{({\bf Q} \cdot {\bf q})^{2}}{{\bf Q}^{2}} \right) \right] \right\}
\label{quattro}
\end{equation}
and
\begin{equation}
  \big< {\bf k} | V_{ \rm{conf} } | {\bf k'} \big>
  = \frac{1}{2 \sqrt{ w_{1} w_{2}
  w_{1}^{\prime} w_{2}^{\prime} } }
  \frac{1}{( 2 \pi )^{3} }
  \int \! d^{3} {\bf r} \, e^{i {\bf Q} \cdot {\bf r} }
  J_{ \rm{si} }({\bf r},{\bf q},q_{10},q_{20})
\label{cinque}
\end{equation}
with
\begin{eqnarray}
  J_{ \rm{si} }({\bf r},{\bf q},q_{10},q_{20}) =
  \frac{ \sigma r }{ q_{10} + q_{20} }
  \bigg[ q_{20}^{2} \sqrt{ q_{10}^{2} - {\bf q}_{ \rm{T} }^{2} } +
  q_{10}^{2} \sqrt{ q_{20}^{2} - {\bf q}_{ \rm{T} }^{2} } +
\nonumber \\
  + \frac{ q_{10}^{2} q_{20}^{2} }{ | {\bf q}_{ \rm{T} } | }
  \bigg( \arcsin \frac{ | {\bf q}_{ \rm{T} } | }{q_{10}} +
  \arcsin \frac{ | {\bf q}_{ \rm{T} } | }{q_{20}} \bigg) \bigg].
\label{sei}
\end{eqnarray}
In this way our hamiltonian becomes strictly related,
but not identical, to that considered in
ref. \cite{tubolsson}.

Let us begin to consider potential
(\ref{quattro}) and (\ref{cinque}) in various different
limit situations. First let us take the
static limit consisting in setting
$ {\bf k} = {\bf k}^{\prime} = 0 $ in
(\ref{quattro}-\ref{sei}).
In this way (\ref{quattro}) and (\ref{cinque})
become local and the hamiltonian can be written as
\begin{equation}
  H_{ \rm{stat} } = \sqrt{ m_{1}^{2} + {\bf k}^{2} } +
    \sqrt{ m_{2}^{2} + {\bf k}^{2} }
    - \frac{4}{3} \frac{ \alpha_{ \rm{s} } }{r} + \sigma r.
\label{sette}
\end{equation}
Less drastically let us assume
$ {\bf q}_{ \rm{T} }^{2} \ll {\bf q}^{2} $
(small angular momentum) and expand (\ref{sei}) in such quantity.
Introducing the regularized Fourier transform
\begin{equation}
  \big< {\bf k} | \sigma r | {\bf k'} \big> =
  - \frac{ \sigma }{ \pi^{2} } \left( \frac{1}{ {\bf Q}^{4} }
  \right)_{ \! \! \rm{reg} } \! \! =
  \frac{ \sigma }{ \pi^{2} } \frac{1}{2}
  \frac{ d^{2} }{ d \varepsilon^{2} }
  \frac{1}{ {\bf Q}^{2} + \varepsilon^{2} }
\end{equation}
(the limit $ \varepsilon \rightarrow 0 $ being understood) and using the
identity
\begin{equation}
  \big< {\bf k} | r {\bf q}_{ \rm{T} }^{2} | {\bf k'} \big> =
  \big< {\bf k} | \frac{ L^{2} + 1 }{r} | {\bf k'} \big>
\end{equation}
$ L^{2} $ being the orbital
angular momentum of the two particles
(notice that classically
$
  {\bf q}_{ \rm{T} }^{2} = {\bf k}_{ \rm{T} }^{2} =
  \frac{ L^{2} }{ r^{2} }
$),
one finds
\begin{eqnarray}
  \big< {\bf k} | V_{ \rm{conf} } | {\bf k'} \big> =
  - \frac{ \sigma }{ 2 \pi^{2} }
  \frac{1}{ \sqrt{ w_{1}^{\prime} w_{2}^{\prime} w_{1} w_{2} } }
  \left[ 2 q_{10} q_{20}
  \left( \frac{1}{ {\bf Q}^{4} } \right)_{ \! \! \rm{reg} } + \right.
\nonumber \\
  + \left. \frac{1}{6} \frac{1}{ q_{10} + q_{20} } \left(
  \frac{ q_{1 0}^{2} }{ q_{20} } +
  \frac{ q_{2 0}^{2} }{ q_{10} } \right)
  \big< {\bf k} | L^{2} + 1 | {\bf k'} \big>
  \frac{1}{ {\bf Q}^{2} } + \cdots \right].
\label{otto}
\end{eqnarray}

On the contrary setting in eq. (\ref{sei}) 
$
{\bf q}_{ \rm{T} }^{2} 
\simeq 
{\bf q}^{2}
$
(large angular momentum) one has
\begin{eqnarray}
  \big< {\bf k} | V_{ \rm{conf} } | {\bf k'} \big> =
  - \frac{ \sigma }{ 2 \pi^{2} }
  \frac{1}{ \sqrt{ w_{1}^{\prime} w_{2}^{\prime} w_{1} w_{2} } }
  \frac{1}{ q_{10} + q_{20} }
  \left( \frac{1}{ {\bf Q}^{4} } \right)_{ \! \! \rm{reg} }
  \left[ q_{2 0}^{2} \sqrt{ q_{1 0}^{2} - {\bf q}^{2} } +
  \right.
\nonumber \\
  + \left. q_{1 0}^{2} \sqrt{ q_{2 0}^{2} - {\bf q}^{2} } +
  \frac{ q_{1 0}^{2} q_{2 0}^{2} }{ | {\bf q} | } \left( \arcsin
  \frac{ | {\bf q} | }{ q_{10} } + \arcsin
  \frac{ | {\bf q} | }{ q_{20} } \right) \right].
\label{nove}
\end{eqnarray}

Finally if in eq.~(\ref{nove}) we consider
the extreme case $ m_{1} = m_{2} = 0 $,
neglect $ V_{ \rm{pert} } $ and simply set 
$ {\bf k} = {\bf k}^{\prime} $ in the factor
multiplying the singular term
\begin{equation}
  \frac{1}{ {\bf Q}^{4} } = \frac{1}{ 
  ( {\bf k'} - {\bf k} )^{4} },
\end{equation}
we obtain the hamiltonian
\begin{equation}
  H_{ \rm{hqt} } = 2 | {\bf q} | + \frac{ \pi }{4} \sigma r
\label{lclcnf}
\end{equation}

In literature
eq.~(\ref{sette}) has been used as such. In ref.~\cite{fulcher}, 
e.g. an overall fit of the low angular momentum
meson spectrum has been obtained by setting
$ \sigma = 0.22 $ GeV$ ^{2} $, $ \alpha_{ \rm{s} } = 0.323 $
and adding to the lefthand of eq.~(\ref{sette}) an {\sl ad hoc}
flavour dependent constant $ C $ in the sector of the
light quarks. As well known, however, for small $ m_{1} $ and $ m_{2} $
eq.~(\ref{sette}) would give asymptotic straight Regge trajectories with
slope $ \alpha^{\prime} = 1/8 \sigma $, which for
$ \sigma = 0.22 $ GeV$ ^{2} $ gives
$ \alpha^{\prime} = 0.57 $ GeV$ ^{- 2} $
while the experimental value
is about $ \alpha^{\prime} = 0.88 $ GeV$ ^{- 2} $.

On the contrary from eq. (\ref{lclcnf}) we would obtain
\begin{equation}
  \alpha^{\prime} = \frac{ 1 }{ 8 \frac{ \pi }{ 4 } \sigma }
   = \frac{ 1 }{ 2 \pi \sigma },
\label{pend}
\end{equation}
which is identical to the Nambu-Goto string model.
This equation gives the correct experimental value for
$ \sigma \simeq 0.18 $ GeV$ ^{2} $, which is the value used
e.g. in ref. \cite{tstrlt1,tstrlt2}
to fit the $ c \bar{c} $ and the $ b \bar{b} $ spectra
(when the coupling with the decay channels is neglected).
This result illustrates the advantage of considering
the potential (\ref{quattro}-\ref{sei}) with respect to a
simply local potential of the type appearing in (\ref{sette}).

Actually
we do not succeed in directly diagonalizing the hamiltonian
eqs. (\ref{eq1a},\ref{quattro}-\ref{sei}), due to problems
of numerical stability and computer time. Therefore
we follow the following strategy.
First we diagonalize the static hamiltonian
$ H_{ \rm{stat} } $ by the Rayleigh-Ritz variational method
using the harmonic oscillator wavefunctions as basis
\cite{lucha96,simon}.
These wavefunctions in coordinate space take the form
\begin{equation}
  \phi_{nlm}( \lambda , {\bf r} ) =
  \Phi_{nl}( \frac{r}{ \lambda } ) Y_{lm}(\hat{r}) =
  \frac{1}{\pi^{ \frac{1}{4} }}
  \frac{1}{\lambda^{ \frac{3}{2} }}
  \sqrt{ \frac{ 2^{n+l+1} (n-1)! }{ [2(n+l)-1]!! } }
  \left( \frac{r}{ \lambda } \right)^{l}
  L_{n-1}^{(l+ \frac{1}{2})}( \frac{ r^{2} }{ \lambda^{2} } )
  e^{- \frac{ r^{2} }{ 2 \lambda^{2} } } Y_{lm}(\hat{r})
\end{equation}
and in the momentum space
\begin{eqnarray}
  \phi_{nlm}( \lambda , {\bf k} ) &=&
  \Phi_{nl}( \lambda k ) Y_{lm}(\hat{k}) = \nonumber \\
  &=& \frac{ (-1)^{n-1} }{\pi^{ \frac{1}{4} }}
  \lambda^{ \frac{3}{2} } ( - i )^{l}
  \sqrt{ \frac{ 2^{n+l+1} (n-1)! }{ [2(n+l)-1]!! } }
  ( \lambda k )^{l}
  L_{n-1}^{(l+ \frac{1}{2})}( \lambda^{2} k^{2} )
  e^{- \frac{ \lambda^{2} k^{2} }{2} } Y_{lm}(\hat{k}),
\label{duetoeq}
\end{eqnarray}
where $ \lambda $ is a scale factor
and $ L_{n-1}^{(l+ \frac{1}{2})} $
are the Laguerre polynomials \cite{weniger}. Then we
write the eigenfunctions of $ H_{\rm stat} $ as
\begin{equation}
  \psi_{nlm}( \lambda , {\bf k} ) = \sum_{n'} a_{n n'}^{l}
  \phi_{n' lm}( \lambda , {\bf k} ) =
  \Psi_{nl}( \lambda k ) Y_{lm}(\hat{k})
\label{atfnz}
\end{equation}
and evaluate the expectation value
\begin{equation}
  \big< \psi_{nlm} | H | \psi_{nlm} \big>.
\label{dieci}
\end{equation}
Finally the eigenvalues of $ H $ are estimated as the minima
of (\ref{dieci}) in $ \lambda $.

Due to eq. (\ref{duetoeq})
the evaluation of the kinetic term in eq. (\ref{dieci}) is
simply reduced to a one dimensional integral.

For the perturbative term, by setting
\begin{equation}
  z_{0} = \frac{ k^{\prime 2} + k^{2} }{ 2 k' k }, \; \; \; \;
  z_{\varepsilon} = \frac{ k^{\prime 2} + k^{2} +
  \varepsilon^{2} }{ 2 k' k }
\end{equation}
we obtain from eq. (\ref{quattro})
\begin{eqnarray}
  \big< \psi_{nlm} | V_{ \rm{pert} } | \psi_{nlm} \big> =
  - \frac{4}{3} \frac{ \alpha_{ \rm{s} } }{ \pi } \frac{1}{2} 
  \int_{0}^{ \infty } \! d k' \, k'
  \frac{1}{ \sqrt{ w_{1}^{\prime} w_{2}^{\prime} } }
  \Psi_{nl}^{\ast}( \lambda k' )
  \int_{0}^{ \infty } \! d k \, k \frac{1}{ \sqrt{ w_{1} w_{2} } }
  \Psi_{nl}( \lambda k )
\nonumber \\
  \left\{ \left( 2 q_{10} q_{20} + \frac{ k^{\prime 2} + k^{2} }{2}
  \right) Q_{l}(z_{0}) + \frac{ k' k }{2l+1} \big[
  (l+1) Q_{l+1}(z_{0}) + l Q_{l-1}(z_{0}) \big]  \right. +
\label{undici}    \\
  \left. + \frac{ ( k' - k )^{2} ( k' + k )^{2} }{4 k' k }
  \left[ \left( \frac{d}{ d z_{0} } P_{l}(z_{0}) \right) \ln \left|
  \frac{ k' + k }{ k' - k } \right| - \frac{d}{ d z_{0} }
  W_{l-1}(z_{0}) \right] - k' k P_{l}(z_{0}) \right\}
\nonumber
\end{eqnarray}
($ P_{l} $ and $ Q_{l} $ being respectively the first and second
kind Legendre polynomials) a double integral.

For what concerns
$ \big< \psi_{nlm} | V_{ \rm{conf} } | \psi_{nlm} \big> $,
if we had used
the original expression (\ref{cinque},\ref{sei}), we should have ended
in a five-dimensional integral which numerical evaluation
is quite problematic for a strongly oscillating function like
$ \psi_{nlm}( \lambda , {\bf k} ) $.
For this reason in place of (\ref{cinque}) and (\ref{sei})
we take in turn advantage of the approximate
eqs. (\ref{otto}) or (\ref{nove}).
From eq. (\ref{otto}) we obtain
\begin{eqnarray}
  \big< \psi_{nlm} | V_{ \rm{conf} } | \psi_{nlm} \big> =
  \frac{ \sigma }{ 2 \pi }
  \int_{0}^{\infty} \! dk' \, k' \frac{1}{
  \sqrt{ w_{1}^{\prime} w_{2}^{\prime} } }
  \Psi_{nl}^{\ast}( \lambda k' ) \int_{0}^{\infty} \! dk \,
  k \frac{1}{ \sqrt{ w_{1} w_{2} } }
  \Psi_{nl}( \lambda k )
\nonumber \\
  \left[ 2 q_{10} q_{20} \lim_{\varepsilon \rightarrow 0}
  \frac{ d^{2} }{ d \varepsilon^{2} }
  Q_{l}( z_{\varepsilon} )
  - \frac{1}{3} \frac{1}{ q_{10} + q_{20} } \left(
  \frac{ q_{1 0}^{2} }{ q_{20} } +
  \frac{ q_{2 0}^{2} }{ q_{10} } \right) ( L^{2} + 1 )
  Q_{l}(z) \right]
\label{dodici}
\end{eqnarray}
(again a double integral)
and from eq. (\ref{nove})
\begin{eqnarray}
 \big< \psi_{nlm} | V_{ \rm{conf} } | \psi_{nlm} \big> &=&
  \frac{ \sigma }{ 2 \pi }
  \int_{0}^{\infty} \! dk' \, k^{\prime 2} \frac{1}{
  \sqrt{ w_{1}^{\prime} w_{2}^{\prime} } }
  \Psi_{nl}^{\ast}( \lambda k' ) \int_{0}^{\infty} \! dk \,
  k^{2} \frac{1}{ \sqrt{ w_{1} w_{2} } }
  \Psi_{nl}( \lambda k ) \frac{1}{ q_{10} + q_{20} } \nonumber \\
& &  \int_{-1}^{1} \!
  d \xi \, P_{l}( \xi )
  \lim_{\varepsilon \rightarrow 0} \frac{ d^{2} }{ d \varepsilon^{2} }
  \frac{1}{ {\bf Q}^{2} + \varepsilon^{2} }
  \bigg[ q_{2 0}^{2} \sqrt{ q_{1 0}^{2} - {\bf q}^{2} } +
  q_{1 0}^{2} \sqrt{ q_{2 0}^{2} - {\bf q}^{2} } + \nonumber \\
& & + \frac{ q_{1 0}^{2} q_{2 0}^{2} }{ | {\bf q} | } \bigg( \arcsin
  \frac{ | {\bf q} | }{ q_{10} } + \arcsin
  \frac{ | {\bf q} | }{ q_{20} } \bigg) \bigg]
\label{tredici}
\end{eqnarray}
(a triple integral).
The singularities occurring in the integrals in eqs.
(\ref{undici}) and (\ref{dodici}) or (\ref{tredici})
can be handled by the method explained in ref. \cite{maung}.

We use eq. (\ref{dodici}) for the evaluation of the heavy
quarkonia spectrum and for the light quarks S state,
eq. (\ref{tredici}) for the Regge trajectories.

Notice that eq. (\ref{dodici}) produces systematically larger
masses. The difference being of the order of few MeV for the
$ b \bar{b} $ system, between 5 and 15 MeV for
$ c \bar{c} $ and progressively larger if light quarks are
involved. The subsequent term in expansion (\ref{dodici})
would contain a $ L^{2} ( L^{2} + 2 ) $ factor and the convergence
of the expansion soon becomes very slow as $ L $ increases.
For this reason we belive eq. (\ref{tredici}) to be preferred
for light quarks already for the P states.

Once that the spin averaged masses are determined we may
evaluate the S-wave hyperfine splitting by the equation
\begin{equation}
  \Delta =
  \big< \psi_{nlm} | V_{3} - V_{1} | \psi_{nlm} \big> =
  \frac{4}{ 3 \pi } \frac{4}{3} \alpha_{\rm s} \delta_{l0}
  \left[ \int_{0}^{ \infty } \! dk \,
  \frac{ k^{2} }{ \sqrt{ w_{1} w_{2} } }
  \Psi_{nl}( \lambda k) \right]^{2}
\label{eqhyp}
\end{equation}
this is obtained taking into consideration only the pure
$
  \mbox{\boldmath $ \sigma $}_{1}
  \cdot
  \mbox{\boldmath $ \sigma $}_{2}
$
term in eq. (\ref{due}).

We have adopted the following parameters:
$ \alpha_{\rm s} = 0.363 $,
$ \sigma = 0.175 $ GeV$^{2} $,
$ m_{c} = 1.40 $ GeV,
$ m_{b} = 4.81 $ GeV,
$ m_{s} = 200 $ MeV,
$ m_{u} = 10 $ MeV;
no {\sl ad hoc} constant $ C $ has been added to the potential.
The first four values have to be compared with those
obtained from heavy quarkonium fits. E.g.
when pair creation effects are neglected (and
after renormalization of the masses by reabsorbing the constant
$ C $) the values used in ref. \cite{tstrlt1}
(table 2, first column) become
$ \alpha_{\rm s} = 0.363 $,
$ \sigma = 0.178 $ GeV$^{2} $,
$ m_{c}^{\prime} = m_{c} + \frac{C}{2} = 1.397 $ GeV,
$ m_{b}^{\prime} = m_{b} + \frac{C}{2} = 4.792 $ GeV,
(cf. also ref. \cite{tubolsson} fig. 4).
On the contrary the light quark masses are
taken as the baricentral values of the current masses:
$ m_{u}^{\rm current} = 2 $ to 8 MeV,
$ m_{d}^{\rm current} = 5 $ to 15 MeV,
$ m_{s}^{\rm current} = 100 $ to 300 MeV
as reported from the Particle Data Group \cite{prtdatb}.
Notice also in this connection
$ m_{c}^{\rm current} = 1.0 $ to 1.6 GeV,
$ m_{b}^{\rm current} = 4.1 $ to 4.5 GeV.
The small rearrangement in the values of $ \sigma $,
$ m_{c} $ and $ m_{b} $ is required to obtain the exact slope
for the Regge trajectories without modifying the
$ \psi $ and the $ \Upsilon $ ground states. Notice that apart
from that no attempt of optimizing the parameter is made.

The results of the calculation for the
$ c \bar{c} $, $ b \bar{b} $
systems are reported in table \ref{tabpes}
and compared with the spin averaged multiplets.
An estimate for such values is taken from ref. \cite{fulcher}.
Also the unperturbed values obtained by $ H_{\rm stat} $ are
reported, as it can be seen the shift for heavy quarks is of
some tenth MeV (but such shift
becomes more important for light quarks).

As expected we have reasonable overall agreement with the data,
the discrepancies being in part
ascribable to pair creation effects \cite{tstrlt1}.

\begin{figure}[htbp!]
  \begin{center}
    \leavevmode
    \setlength{\unitlength}{1.0mm}
    \begin{picture}(140,70)
      \put(25,0){\mbox{\epsfig{file=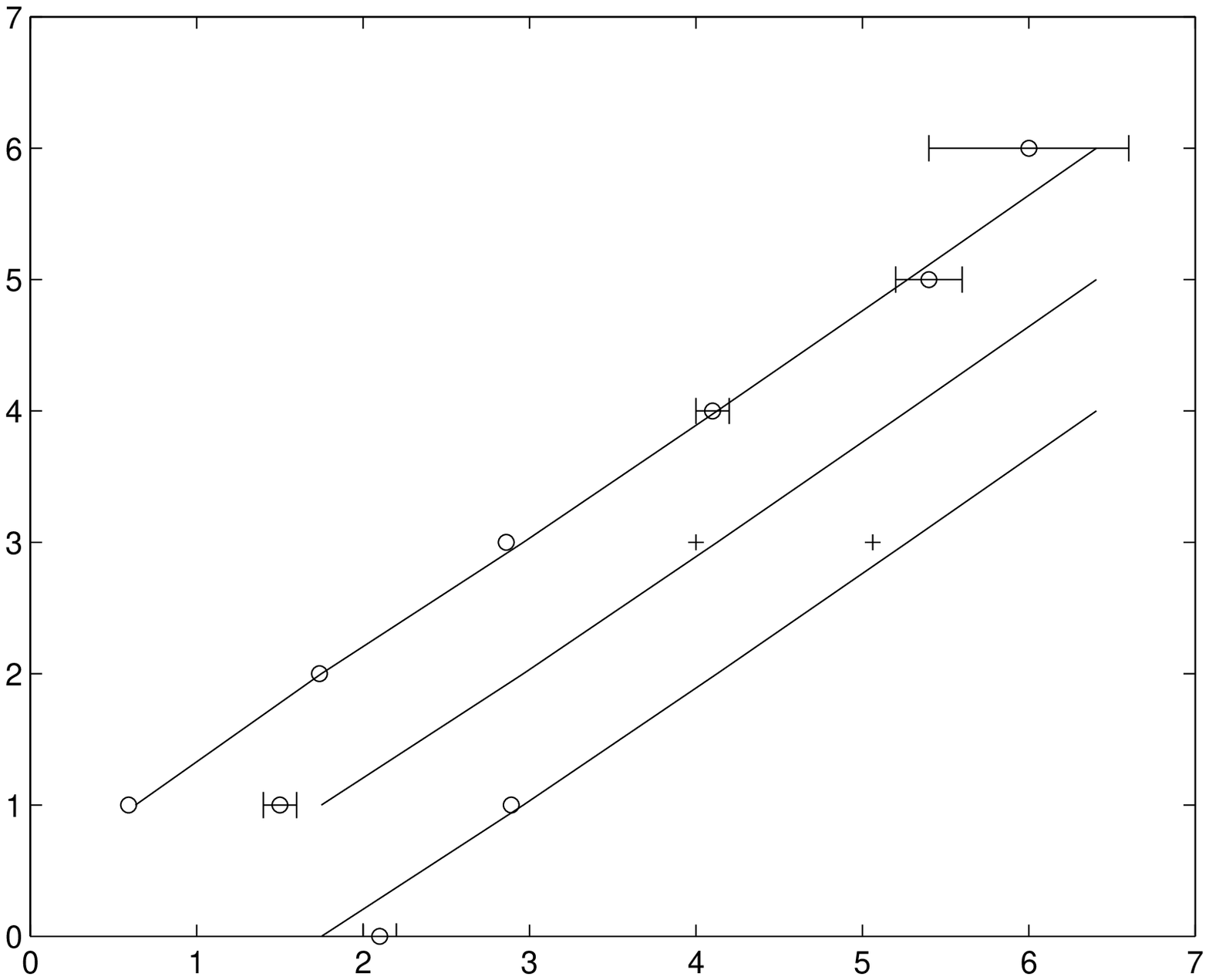,height=7cm}}}
      \put(15,60){ $ J $ }
      \put(110,59){ $ J = L + 1 $ }
      \put(110,49){ $ J = L $ }
      \put(110,39){ $ J = L - 1 $ }
      \put(45,60){ $ \alpha^{ \prime } = 0.878 $ }
      \put(100,-5){ $ M^{2} $ }
      \put(20,14){ $ \rho ( 770 ) $ }
      \put(28,22){ $ a_{2} ( 1320 ) $ }
      \put(32,7){ $ a_{1} ( 1260 ) $ }
      \put(40,-5){ $ a_{0} ( 1450 ) $ }
      \put(41,30){ $ \rho_{3} ( 1690 ) $ }
      \put(59,9){ $ \rho ( 1700 ) $ }
      \put(56,40){ $ a_{4} ( 2040 ) $ }
      \put(61,26){ $ X ( 2000 ) $ }
      \put(70,50){ $ \rho_{5} ( 2350 ) $ }
      \put(84,26){ $ \rho_{3} ( 2250 ) $ }
      \put(90,62){ $ a_{6} ( 2450 ) $ }
    \end{picture}
  \end{center}
\vspace{-0.2cm}
\caption{}
\label{fig1}
\end{figure}

\begin{figure}[htbp!]
  \begin{center}
    \leavevmode
    \setlength{\unitlength}{1.0mm}
    \begin{picture}(140,70)
      \put(25,0){\mbox{\epsfig{file=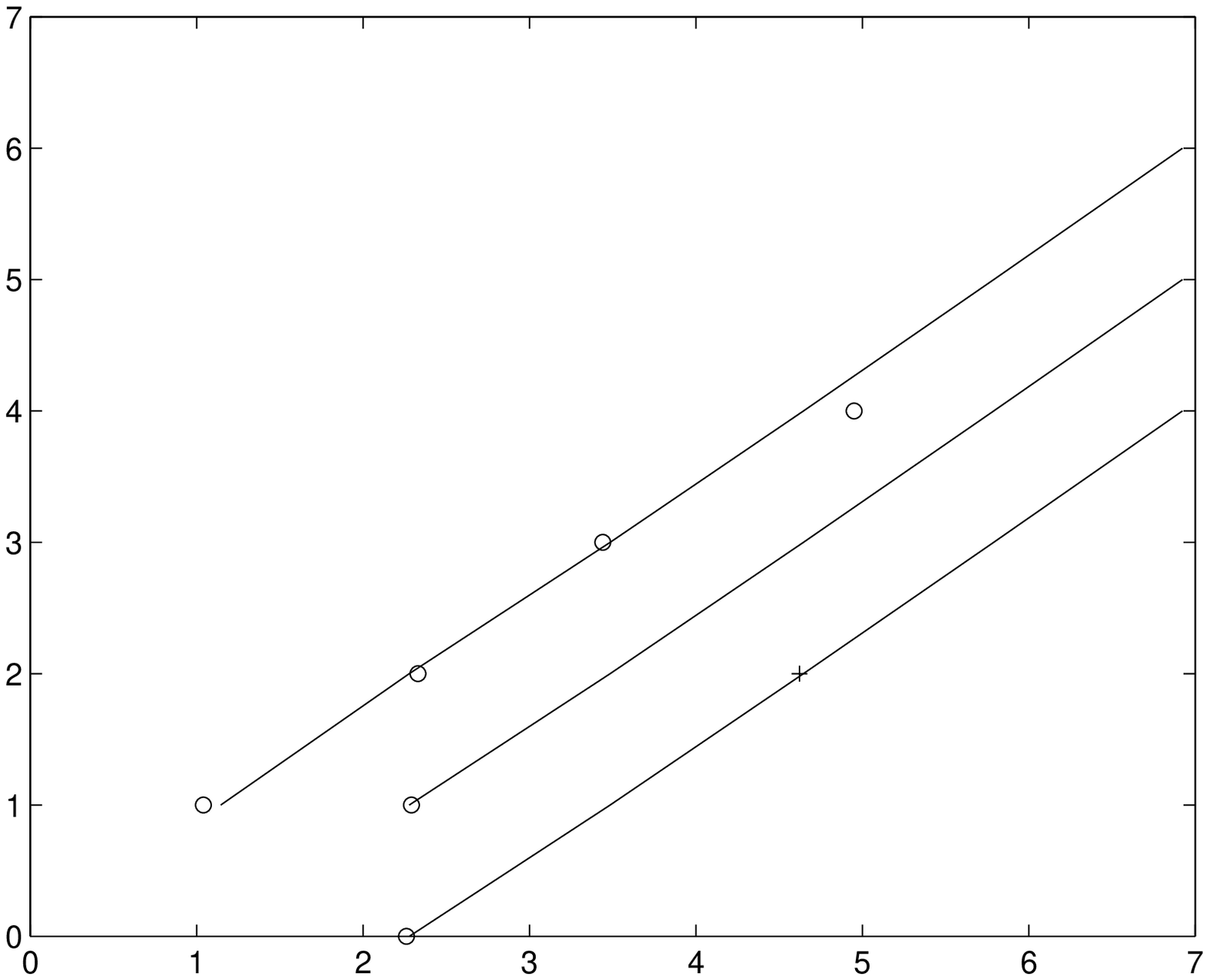,height=7cm}}}
      \put(15,60){ $ J $ }
      \put(110,59){ $ J = L + 1 $ }
      \put(110,49){ $ J = L $ }
      \put(110,39){ $ J = L - 1 $ }
      \put(45,60){ $ \alpha^{ \prime } = 0.878 $ }
      \put(100,-5){ $ M^{2} $ }
      \put(23,14){ $ \phi ( 1020 ) $ }
      \put(35,22){ $ f_{2}^{\prime} ( 1525 ) $ }
      \put(40,8){ $ f_{1} ( 1510 ) $ }
      \put(45,-5){ $ f_{0} ( 1500 ) $ }
      \put(47,31){ $ \phi_{3} ( 1850 ) $ }
      \put(65,43){ $ f_{j} ( 2220 ) $ }
      \put(81,18){ $ f_{2} ( 2150 ) $ }
    \end{picture}
  \end{center}
\vspace{-0.2cm}
\caption{}
\label{fig2}
\end{figure}

\begin{figure}[htbp!]
  \begin{center}
    \leavevmode
    \setlength{\unitlength}{1.0mm}
    \begin{picture}(140,70)
      \put(25,0){\mbox{\epsfig{file=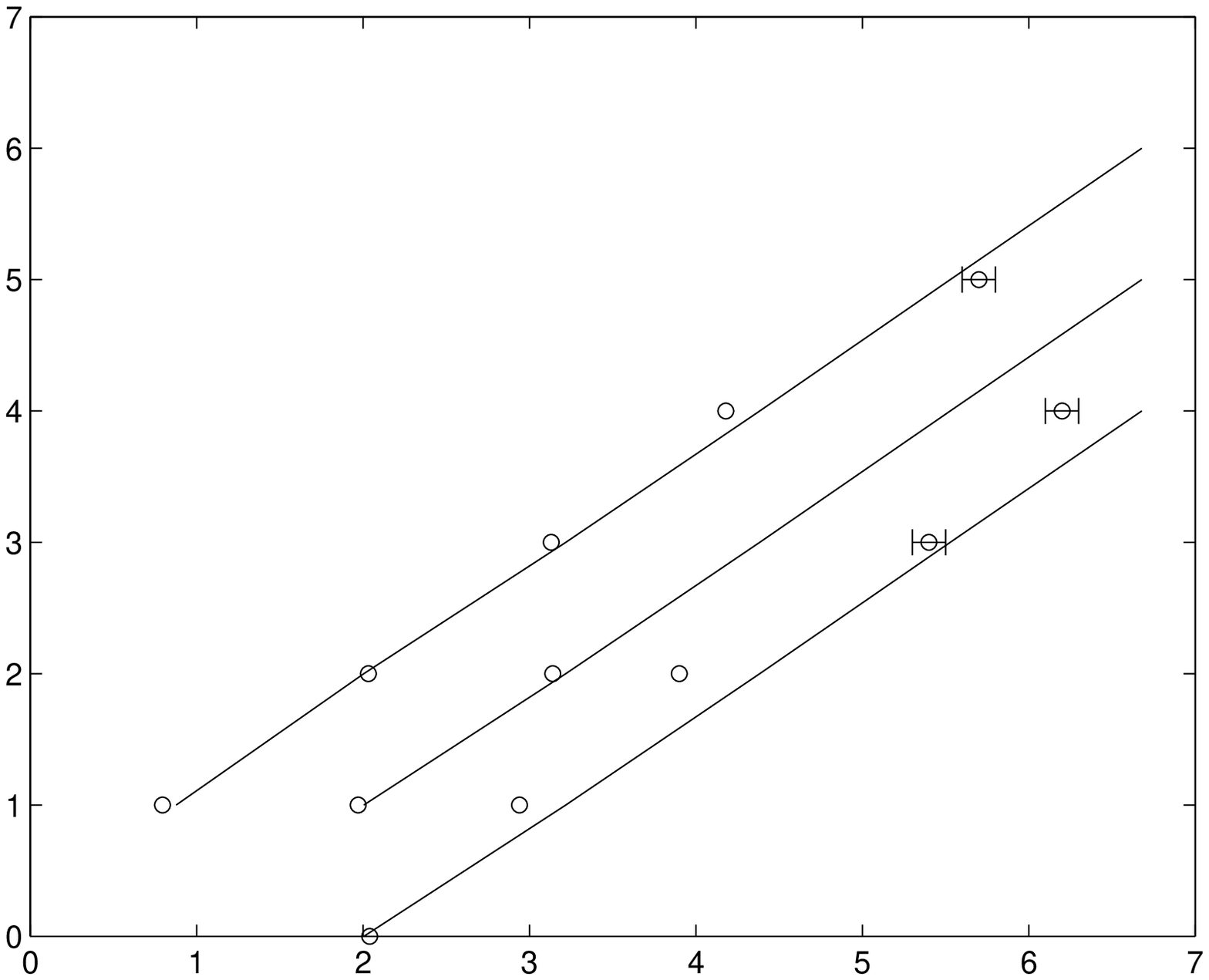,height=7cm}}}
      \put(15,60){ $ J $ }
      \put(110,59){ $ J = L + 1 $ }
      \put(110,49){ $ J = L $ }
      \put(110,39){ $ J = L - 1 $ }
      \put(45,60){ $ \alpha^{ \prime } = 0.874 $ }
      \put(100,-5){ $ M^{2} $ }
      \put(22,14){ $ K^{\ast} ( 892 ) $ }
      \put(30,22){ $ K_{2}^{\ast} ( 1430 ) $ }
      \put(35,8){ $ K_{1} ( 1400 ) $ }
      \put(40,-3){ $ K_{0}^{\ast} ( 1430 ) $ }
      \put(44,31){ $ K_{3}^{\ast} ( 1780 ) $ }
      \put(50,17){ $ K_{2} ( 1770 ) $ }
      \put(63,9){ $ K^{\ast} ( 1680 ) $ }
      \put(56,40){ $ K_{4}^{\ast} ( 2045 ) $ }
      \put(84,30){ -- }
      \put(78.5,30){ $ \longleftarrow $ }
      \put(90,28){ $ K_{3} ( 2320 ) $ }
      \put(72,21){ $ \rightarrow $ }
      \put(77,19){ $ K_{2}^{\ast} ( 1980 ) $ }
      \put(72,50){ $ K_{5}^{\ast} ( 2380 ) $ }
      \put(92,39.5){ $ \leftarrow $ }
      \put(90,43){ $ K_{4} ( 2500 ) $ }
    \end{picture}
  \end{center}
\vspace{-0.2cm}
\caption{}
\label{fig3}
\end{figure}

In Fig. \ref{fig1}, \ref{fig2}, \ref{fig3} the Regge trajectories
are reported for the ground triplet states of the systems
$ u \bar{u} $, $ s \bar{s} $, $ u \bar{s} $ corresponding to
$ J = L + 1 , \, L , \, L - 1 $. The agreement is again
usually very good, particularly for the states $ J = L + 1 $,
while some disagreement, particularly in the $ u \bar{s} $ case,
can possibly be retraced in the neglecting of the spin-orbit terms.
In table \ref{tableg1} numerical results are reported
for the $ u \bar{u} $ system and compared
with the data for the $ J = L + 1 $ state. The values in
bracket for the 1S and 2S states are obtained by adding 1/4 of the
hyperfine splitting as given in table \ref{tableg8}.
Notice that the 2S state of table \ref{tableg1} is too high by about
100 MeV, such discrepancy is however smaller than the width of
$ \rho ( 1450 ) $ and $ \omega ( 1420 ) $ and again could be
ascribable to pair creation effects.
Similar circumstances occur for the $ u \bar{s} $ and
$ s \bar{s} $ cases.

The results for $ u \bar{c} $, $ u \bar{b} $,
$ s \bar{c} $, and $ s \bar{b} $ systems
are reported in table \ref{tablegn} and compared with the
experimental spin averaged masses using the theoretical
splitting when the singlet state has not yet been observed.

In table \ref{tableg8} the hyperfine splittings as
evaluated by eq. (\ref{eqhyp}) are reported for the
heavy-heavy and the heavy-light cases.
In the light-light case
a comparison with the splitting $ \rho $--$ \pi $,
$ \varphi $--$ \eta^{\prime} $ and $ K^{\ast} $--$ K $
would have no meaning for the reason we have explained.
Notice however that for the $ u \bar{u} $ ground state
eq. (\ref{eqhyp}) would give $ \Delta = 221 $ MeV in good
agreement with the $ \omega $--$ \eta $ splitting (234 MeV).

The fact that we do not have to use constituent masses in our
calculation seem surprising at first sight. Notice
however that, following \cite{martin},
if we expand the kinetic part of the hamiltonian
eq. (\ref{eq1a}) around the expectation value
$ \big< \psi_{nlm} | {\bf k}^{2} | \psi_{nlm} \big> $
(that we call simply $ \big< {\bf k}^{2} \big> $)
we find
\begin{equation}
  M = m_{1 {\rm eff} } + m_{2 {\rm eff} } +
  \frac{ {\bf k}^{2} }{ 2 m_{1 {\rm eff} } } +
  \frac{ {\bf k}^{2} }{ 2 m_{2 {\rm eff} } }
  - \frac{ {\bf k}^{4} }{ 8 m_{1 {\rm eff} }^{3} }
  - \frac{ {\bf k}^{4} }{ 8 m_{2 {\rm eff} }^{3} }
  + V + C
\end{equation}
where at this order
\begin{equation}
  m_{i \, {\rm eff} } = 
  \sqrt{ m_{i}^{2} + \big< {\bf k}^{2} \big> }
  \left( 1 + \frac{1}{2} \frac{ \big< {\bf k}^{2}
  \big> }{ m_{i}^{2} + \big< {\bf k}^{2} \big> }
  \right)^{\! \! - 1},
\hspace{1cm}
  i = 1,2
\end{equation}
and the quantity $ C $ can be defined by difference.

Average values of
$
  m_{u \, {\rm eff} }
$
and
$
  m_{s \, {\rm eff} }
$
for the lowest states involving quarks $ u $ and $ s $
obtained evaluating $ \big< {\bf k}^{2} \big> $ for the wave
function eq. (\ref{atfnz}) are typically
$
  m_{u \, {\rm eff} } = 360
$ MeV,
$
  m_{s \, {\rm eff} } = 410
$ MeV,
$
  m_{c \, {\rm eff} } = 1.440
$ GeV,
$
  m_{b \, {\rm eff} } = 4.820
$ GeV, with
$
  C \simeq 300
$ MeV.
These are of the same order of the constituent masses
used in semirelativistic computations \cite{martin}.

In conclusion, starting from our first principle Salpeter
equation, we have obtained an overall good reproduction of the
spectrum of the mesons involving heavy and light quarks with
the exception of the light pseudoscalar states. Since the
parameter are practically completely specified by the heavy
quarkonium spectrum and by high energy scattering,
our calculation is essentially parameter free,
for what concerns light-light and light-heavy quark systems.

We would like to thank K. Maung Maung, L. Sorrillo and R. Chen
for the useful discussions and N. Brambilla and A. Vairo
for assistance on numerical problems.
\begin{table}
\centering
\caption{Spin averaged energies for heavy quarkonium systems.
Our theoretical results and experimental values.}
\begin{tabular}{cccl}
\hline
States & static potential & total potential & experimental values \\
              &    (GeV)  &    (GeV)  &      (GeV)           \\
\hline
$ b \bar{b} $ &        &           &                      \\
   1S      &  9.504    &   9.446   & ~9.448 $ \pm $ 0.005 \\
   2S      & 10.055    &  10.014   & 10.017 $ \pm $ 0.005 \\
   3S      & 10.385    &  10.347   & 10.351 $ \pm $ 0.005 \\
   4S      & 10.652    &  10.615   & 10.580 $ \pm $  ~~?  \\
   5S      & 10.886    &  10.849   & 10.865 $ \pm $  ~~?  \\
   6S      & 11.109    &  11.068   & 11.019 $ \pm $  ~~?  \\
   1P      &  9.972    &   9.953   & ~9.900 $ \pm $ 0.001 \\
   2P      & 10.310    &  10.288   & 10.260 $ \pm $ 0.001 \\
\hline
$ c \bar{c} $ &        &           &                      \\
   1S      &  3.133    &   3.065   &  3.067 $ \pm $ 0.002 \\
   2S      &  3.693    &   3.623   &  3.663 $ \pm $ 0.005 \\
   3S      &  4.100    &   4.030   &  4.040 $ \pm $ ~~?   \\
   4S      &  4.441    &   4.372   &  4.415 $ \pm $ ~~?   \\
   1P      &  3.551    &   3.508   &  3.525 $ \pm $ 0.001 \\
   1D      &  3.833    &   3.794   &  3.770 $ \pm $ ~~?   \\
   2D      &  4.201    &   4.158   &  4.159 $ \pm $ ~~?   \\
\hline
\end{tabular}
\label{tabpes}
\end{table}
\begin{table}
\centering
\caption{ Theoretical results and experimental data for $ u \bar{u} $
system. For the S states we write the theoretical averaged
mass and in brackets the mass for the triplet state obtained
with the hyperfine splitting collected in table \ref{tableg8}. }
\begin{tabular}{cccc}
\hline
 State & M theor. & \multicolumn{2}{c} {Exper. data (GeV)} \\
\cline{3-4}
 & (GeV) & $ \omega \; \; \; I^{G}(J^{PC}) $ &
 $ \rho \; \; \; I^{G}(J^{PC}) $
 \\
\hline
$ 1 \, {^{3} {\rm S}_{1}} $ &
 0.740 & $ \omega (782) \; 0^{-} (1^{- -}) $ &
$ \rho (770) \; 1^{+} (1^{- -}) $
 \\
 & (0.795) & $ 0.78194 \pm 0.00012 $ & $ 0.7685 \pm 0.0006 $
 \\
\hline
$ 2 \, {^{3} {\rm S}_{1}} $ &
 1.543 & $ \omega (1420) \; 0^{-} (1^{- -}) $ &
$ \rho (1450) \; 1^{+} (1^{- -}) $
 \\
 & (1.556) & $ 1.419 \pm 0.031 $ & $ 1.465 \pm 0.025 $
 \\
\hline
$ 1 \, {^{3} {\rm P}_{2}} $ &
 1.323 & $ f_{2} (1270) \; 0^{+} (2^{+ +}) $ &
$ a_{2} (1320) \; 1^{-} (2^{+ +}) $
 \\
 &  & $ 1.275 \pm 0.005 $ & $ 1.3181 \pm 0.0007 $
 \\
\hline
$ 1 \, {^{3} {\rm D}_{3}} $ &
 1.721 & $ \omega_{3} (1667) \; 0^{-} (3^{- -}) $ &
$ \rho_{3} (1690) \; 1^{+} (3^{- -}) $
 \\
 &  & $ 1.667 \pm 0.004 $ & $ 1.691 \pm 0.005 $
 \\
\hline
$ 1 \, {^{3} {\rm F}_{4}} $ &
 2.031 & $ f_{4} (2050) \; 0^{+} (4^{+ +}) $ &
$ a_{4} (2040) \; 1^{-} (4^{+ +}) $
 \\
 &  & $ 2.044 \pm 0.011 $ & $ 2.037 \pm 0.026 $
 \\
\hline
$ 1 \, {^{3} {\rm G}_{5}} $ &
 2.296 &  & $ \rho_{5} (2350) \; 1^{+} (5^{- -}) $
 \\
 &  &  & $ 2.330 \pm 0.035 $
 \\
\hline
$ 1 \, {^{3} {\rm H}_{6}} $ &
 2.531 & $ f_{6} (2510) \; 0^{+} (6^{+ +}) $ &
$ a_{6} (2450) \; 1^{-} (6^{+ +}) $
 \\
 &  & $ 2.510 \pm 0.030 $ & $ 2.450 \pm 0.130 $
 \\
\hline
\end{tabular}
\label{tableg1}
\end{table}
\begin{table}
\caption{Theoretical results for
$ u \bar{c} $, $ u \bar{b} $, $ s \bar{c} $, $ s \bar{b} $ systems
(MeV). Experimental data are enclosed in brackets.}
\begin{tabular}{ccccc}
\hline
 State & $ u \bar{c} $ & $ u \bar{b} $ &
 $ s \bar{c} $ & $ s \bar{b} $ \\
\hline
 1S &
 1973 ($ 1973 \pm 1 $) &
 5326 ($ 5313 \pm 2 $) &
 2080 ($ 2076.4 \pm 0.5 $) &
 5418 ($ 5404.6 \pm 2.5) $
\\
 2S &
 2600 $ ( 2623 \pm ? )^{\rm a} $ &
 5906 $ ( 5897 \pm ? )^{\rm a} $ &
        &
\\
 1P &
 2442 $ ( 2438 \pm ? )^{\rm b} $ &
 5777 $ ( 5825 \pm 14 )^{\rm c} $ &
 2528 ($ 2535.35 \pm 0.34 $)  &
 5848 ($ 5853 \pm 15 $)
\\
\hline
\end{tabular}
\label{tablegn}
$ ^{\rm a} $Obtained from preliminary {\sl Delphi} data
$ m(D^{\ast \prime}) = 2637 \pm 8 $ MeV,
$ m(B^{\ast \prime}) = 5906 \pm 14 $ MeV \cite{pullia}
subtracting 1/4 theoretical hyperfine splitting reported in
table \ref{tableg8}.
\\
$ ^{\rm b} $Estimated from
$ m(D_{2}^{\ast}) = 2459 \pm 4 $ MeV,
$ m(D_{1}) = 2427 \pm 5 $ MeV.
\\
$ ^{\rm c} $From preliminary {\sl Delphi} data \cite{pullia}.
\end{table}
\begin{table}
\centering
\caption{Theoretical results for 
$ q \bar{q} $ hyperfine splitting (MeV).
Experimental data are enclosed in brackets.}
\begin{tabular}{ccccccc}
\hline
 State & $ u \bar{c} $ & $ u \bar{b} $ & $ c \bar{c} $ &
 $ b \bar{b} $ & $ s \bar{c} $ & $ s \bar{b} $ \\
\hline
 1S &
 111 ($ 141 \pm 1 $) &
 59 ($ 46 \pm 3 $) &
 97 ($ 117 \pm 2 $) &
 102 &
 108 (144) &
 60 ($ 47 \pm 4) $    \\
 2S &
 59 &
 38 &
 59 ($ 92 \pm 5 $) &
 42 &
 62 &
 40 \\
\hline
\end{tabular}
\label{tableg8}
\end{table}
\clearpage

\hskip 5 cm {\bf FIGURE CAPTIONS}
\vskip 1 cm
 
\noindent {\bf Fig. 1}\\
\noindent
Ground triplet $ u \bar{u} $ Regge trajectories.
Theoretical results (full line) compared with experimental data
(circlet). Cross denote less established masses.
\vspace{4mm}

\noindent {\bf Fig. 2}\\
\noindent
Ground triplet $ s \bar{s} $ Regge trajectories
(with the same notations of fig. 1).
\vspace{4mm}

\noindent {\bf Fig. 3}\\
\noindent
Ground triplet $ u \bar{s} $ Regge trajectories
(with the same notations of fig. 1).
\vspace{4mm}

\clearpage
\vspace{1.cm}

\end{document}